\documentstyle[emulateapj,danonecolfloat,epsfig]{article}

\def\NoApjSectionMarkInTitle#1{#1.\ }
\newcommand\lsim{\mathrel{\rlap{\lower4pt\hbox{\hskip1pt$\sim$}}
        \raise1pt\hbox{$<$}}}
\newcommand\gsim{\mathrel{\rlap{\lower4pt\hbox{\hskip1pt$\sim$}}
        \raise1pt\hbox{$>$}}}

\begin{document}
\twocolumn[
 
 
\journalid{337}{15 January 1989}
\articleid{11}{14}
 
\submitted{\today. To be submitted to ApJ.}

\title{The Luminosity Dependence of Quasar Clustering}

\author{Adam Lidz$^{1}$, Philip F. Hopkins$^{1}$, Thomas J. Cox$^{1}$, Lars Hernquist$^{1}$, Brant Robertson$^{1}$}
\affil{$^1$ Harvard-Smithsonian Center for Astrophysics, 60 Garden Street, Cambridge, MA 02138, USA}
\begin{abstract}

We investigate the luminosity dependence of quasar clustering,
inspired by numerical simulations of galaxy mergers that incorporate
black hole growth.  These simulations have motivated a new
interpretation of the quasar luminosity function.  In this picture,
the bright end of the quasar luminosity function consists of quasars
radiating nearly at their peak luminosities, while the faint end
consists mainly of very similar sources, but at dimmer phases in their
evolution. We combine this model with the statistics of dark matter
halos that host quasar activity. We find that, since bright and faint
quasars are mostly similar sources seen in different evolutionary
stages, a broad range in quasar luminosities corresponds to only a
narrow range in the masses of quasar host halos.  On average, bright
and faint quasars reside in similar host halos.  Consequently, we
argue that quasar clustering should depend only weakly on luminosity.
This prediction is in qualitative agreement with recent measurements
of the luminosity dependence of the quasar correlation function (Croom
et al. 2005) and the galaxy-quasar cross-correlation function
(Adelberger \& Steidel 2005).  Future precision clustering
measurements from SDSS and 2dF, spanning a large range in luminosity,
should provide a strong test of our model.

\end{abstract}


]

\section{Introduction}
\label{intro}

Recently, black hole growth and feedback have been incorporated into
numerical simulations of galaxy mergers (Springel et al. 2005a,b). In
these simulations, gravitational torques drive inflows of gas into the
nuclei of merging galaxies (e.g. Barnes \& Hernquist 1991, 1996),
triggering starbursts (e.g. Mihos \& Hernquist 1996) and feeding the
growth of central supermassive black holes (Springel et al. 2005a; Di
Matteo et al. 2005).  As the black holes accrete, some of the radiated
energy couples to the surrounding gas, and the growth eventually
stalls when this feedback energy is sufficient to unbind the
surrounding reservoir of gas.  These simulations elucidate the
connection between galaxy evolution, the formation of supermassive
black holes, and the self-regulated nature of quasar activity, and
provide quantitative predictions which agree well with observations
of, e.g., the $M_{\rm BH} - \sigma$ relation (Di Matteo et al. 2005,
Robertson et al. 2005), quasar lifetimes (Hopkins et al. 2005a,b), and
the quasar luminosity function in various wavebands (Hopkins et
al. 2005c,d,e).  One important product of these numerical models is a
quantitative prediction of the light curve of quasar activity and the
resulting quasar ``lifetime'' -- i.e., the amount of time that a
quasar spends at a given luminosity -- and its dependence on the
properties of the merging galaxies.

The simulated quasar light curves imply a qualitatively different
picture of the quasar luminosity function than previously
considered. Specifically, in the model of Hopkins et al. (2005a-e),
the bright end of the quasar luminosity function consists of systems
radiating at close to their peak luminosities. The faint end of the
luminosity function is dominated by similar quasars, observed,
however, in a faint stage of their life cycle.  This occurs because
the quasar lifetime in this model is longer at lower luminosities;
i.e.\ a given quasar spends more time (and is more likely to be
observed) at a luminosity well below its peak luminosity. This differs
from previous models which generally assume that quasars radiate at a
fixed luminosity for some characteristic lifetime, or adopt simplified
exponential light curves (e.g., Kauffmann \& Haehnelt 2000, Wyithe \& Loeb 2003).

One way to distinguish our picture from previous models is through its
predictions for the luminosity dependence of quasar clustering. In
most semi-analytic models to date, there is assumed to be a tight
relation between the {\em instantaneous} luminosity of a quasar and
the mass of its host halo.  In these analyses, faint quasars populate
low mass halos, and are less clustered than bright quasars which
populate high mass halos. In the model of Hopkins et al. (2005a-e)
most bright and faint quasars are similar sources, seen at a different stage
in their evolution.  Therefore, one expects quasar clustering {\em to
depend less strongly on luminosity} in this scenario than in most
previous models.  Although Hopkins et al.\ (2005e) demonstrate that
this theory reproduces well a wide range of quasar observations not
explained by previous models of the quasar light curve, the luminosity
dependence of quasar clustering provides a direct probe of the most
fundamental distinction between these models, and can be tested even
at high redshift where observations of e.g.\ the Eddington ratio
distribution of quasars are not currently practical.

In fact, Adelberger \& Steidel (2005) have recently measured the
galaxy-quasar cross correlation function, finding no evidence for
luminosity-dependent clustering. Their interpretation of this
observation is that faint quasars are longer-lived than bright
quasars. While this interpretation is qualitatively similar to the one
we advocate, we will further demonstrate that it is, in fact, a
natural consequence of our numerical models.

The aim of the present paper is to provide quantitative predictions
for the luminosity dependence of quasar clustering, based on our
numerical simulations.  Our analysis proceeds in two steps.  The
first, described in \S \ref{lum_mass}, characterizes the relationship
between quasar luminosity and the mass of quasar host dark matter
halos. We use the numerical simulations of Springel et al. (2005a) to
formulate our description of the connection between quasar luminosity
and halo mass. The next step of our analysis is to connect quasar
properties with the statistics of the dark matter halos that host
quasars, as has been done previously (e.g, Efstathiou \& Rees 1998;
Kauffmann \& Haehnelt 2000; Martini \& Weinberg 2001; Haiman \& Hui
2001; Porciani et al. 2004; Wyithe \& Loeb 2005).  This part of our
calculation is described in \S \ref{hosts}, where we determine which
dark matter halos host active quasars, and provide predictions for the
luminosity dependence of quasar bias. In \S \ref{zev} we explore the
redshift evolution of quasar clustering. In \S \ref{conclusion} we
conclude and summarize the present status of, and future prospects
for, measurements of the luminosity dependence of quasar clustering.

\section{The relation between quasar luminosity and halo mass}
\label{lum_mass}

We begin by connecting quasar luminosity with the mass of quasar host
halos. The key point here is that the halo mass is {\em correlated
with the peak luminosity of quasar sources}, and is connected only
indirectly with the {\em instantaneous luminosity} through the quasar
light curve.  We therefore separate the connection between {\em
instantaneous} quasar luminosity and halo mass into two distinct
pieces. The first part is the correlation between peak luminosity and
halo mass. The second part involves the quasar light curve which
connects the instantaneous and peak luminosities of quasar activity.
We illustrate this by evaluating the correlation between halo mass and
peak luminosity, considering numerical simulations of merging galaxies
at $z = 2$, i.e., close to the epoch of peak quasar activity.

\begin{figure}[t]
\vbox{ \centerline{
\plotone{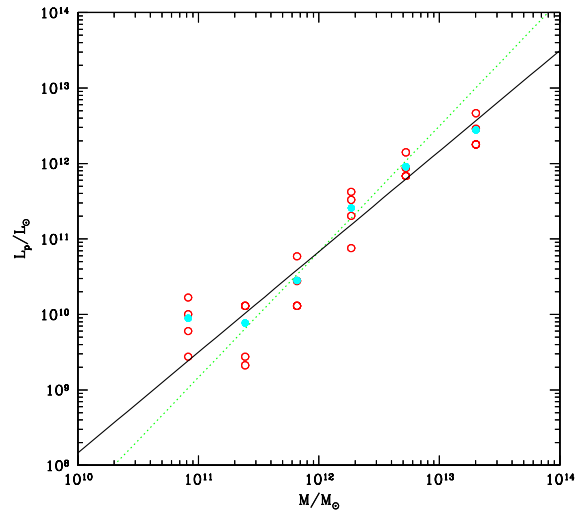}}
\caption[f1.eps]{Correlation between peak B-band quasar luminosity and halo mass at $z=2$.
The open red circles indicate the maximum, or peak B-band luminosity of
quasar activity for each of $24$ simulated halos at $z=2$ as a function of the mass of
the quasar's host halo. The closed cyan circles indicate the mean B-band luminosity for quasars
of a given halo mass. The green dotted line indicates the correlation expected from the model
of Wyithe \& Loeb (2003). The black solid line indicates the correlation expected from
analytic models of momentum-driven outflows (e.g., King 2003).
\label{peakl_v_mass}} }
\end{figure} 

Specifically, we consider $24$ simulations of merging galaxies at
$z=2$.  The simulations, described in more detail in Springel et
al. (2005a) and Robertson et al. (2005), model merging galaxies of
varying halo mass, initial disk gas fraction, and effective equation
of state for the interstellar gas. In each simulation, we determine
the amount of time the merger spends in each of several logarithmic
bins in bolometric luminosity. We then determine the peak bolometric
luminosity of a particular merger by identifying the highest
luminosity bin reached.  The peak bolometric luminosity is converted
into a peak optical, B-band luminosity using the relation of Marconi
et al. (2004), given by log$_{\rm 10} \left( L_B
\right) = 0.80 - 0.067 {\cal{L}} + 0.017 {\cal{L}}^2 - 0.0023
{\cal{L}}^3$, with ${\cal{L}} = $ log$_{\rm 10} \left(L_{\rm
bol}/L_\odot \right) - 12$. Here, $L_B$ denotes the quasar B-band
luminosity, while $L_{\rm bol}$ indicates the quasar bolometric
luminosity.\footnote{In the remainder of the paper, we will use $L$ to
denote the optical, B-band luminosity of quasar activity -- i.e, we
will generally suppress the subscript `B'.}  The resulting peak B-band
luminosity from our simulations, and its dependence on halo mass is
shown in Fig. \ref{peakl_v_mass}. The red open circles in the plot
indicate the result of each merger simulation, while the cyan closed
circles show the results (logarithmically) averaged over simulations with identical
halo mass.

The plot indicates a clear correlation between the peak luminosities
of quasars
and the masses of their host dark matter halos. This correlation is a
natural consequence of the self-regulated nature of quasar activity.
Specifically, analytic models of self-regulated black hole growth predict
that the peak luminosity scales with the halo circular velocity, $v_c$,
either as $L_p \propto v_c^5$, or as $L_p \propto v_c^4$. The first
scaling is the result of assuming that the peak luminosity is set by
equating the feedback energy from accretion 
coupled to the halo gas in a dynamical time with
the binding energy of the gas (Silk \& Rees 1998; Ciotti \& Ostriker
2001; Wyithe \& Loeb 2002). The second scaling results from assuming that
momentum, as opposed to energy, is conserved in the evolution of the quasar
``outflow'' that eventually unbinds the surrounding gas (e.g., King 2003, Di Matteo et al.\ 2005). This
scaling is appropriate if the outflowing gas can cool efficiently. 

Indeed, these authors suggest that this self-regulation likely
explains the tight correlation observed between black hole mass and
stellar velocity dispersion in local black hole populations
(e.g. Gebhardt et al. 2000; Ferrarese \& Merritt 2000; Tremaine et
al. 2002), with the peak luminosity corresponding to the Eddington
luminosity of the final black hole mass.  Ferrarese (2002) further
argues that there is direct observational support for a correlation
between black hole mass and halo circular velocity, as expected in
the analytic models.  Since the halo
circular velocity is proportional to the one-third power of the halo
mass, the first scaling implies that peak luminosity scales with
halo mass as $L_p \propto M^{5/3}$, while the second scaling implies
$L_p \propto M^{4/3}$. 

A comparison between these scalings and the simulation results is shown in Fig.
\ref{peakl_v_mass}.
The green dotted line 
indicates the first scaling, with the normalization adopted by Wyithe \&
Loeb (2003) which, at $z = 2$, is $L_p = 6.78
\times 10^{10} L_\odot \left(M/10^{12} M_\odot \right)^{5/3}$. The
second scaling, $L_p \propto M^{4/3}$, with the same normalization, 
is indicated by the black solid
line in the figure. Neither scaling is a perfect description of the mean trend
seen in the simulations, although the somewhat shallower relation, $L_p \propto M^{4/3}$, is 
clearly a better overall match. This corresponds
to a $M_{\rm BH} - \sigma$ relation of $M_{\rm BH} \propto \sigma^4$, (see 
Di Matteo et al. 2005, Robertson et al. 2005 for direct measurements of the 
$M_{\rm BH} - \sigma$ relation in our simulations), 
which agrees better with observations (e.g. Tremaine et al. 2002) than 
the alternate scaling. In practice, we find that halos with only a narrow range in mass
host active quasars (see \S \ref{hosts}). Hence, we find that our results are quite
similar if we use a direct spline fit to the mean simulated trend, (shown by the
cyan circles in Fig. \ref{peakl_v_mass}), or instead adopt the approximate 
$L_p \propto M^{4/3}$ scaling. Likewise, we have also analyzed in detail the difference between 
assuming an $L_p \propto M^{5/3}$ or $L_p \propto M^{4/3}$ scaling, 
and find that they give qualitatively identical results in our subsequent
analysis. 
For simplicity, we therefore adopt the $L_p \propto M^{4/3}$
scaling in what follows.

In addition to the mean correlation between peak luminosity and halo
mass, we would like to incorporate the amount of {\em scatter} in this
relation into our modeling. It is clear from Fig. \ref{peakl_v_mass}
that the level of scatter in our simulations is significant.
Specifically, we find that the average dispersion in the peak
luminosity at fixed halo mass is $\sigma_{\rm L_p}/L_p \sim 0.8$.  We
account for this scatter, and the mean correlation between peak
luminosity and halo mass, by adopting a lognormal form for the
conditional probability distribution of peak luminosity given the
halo mass,
\begin{equation}
\frac{dP(L_p|M)}{dL_p} = \frac{1}{\sqrt{2 \pi \Delta^2} L_p} \rm{Exp}
\left[-\frac{\left(\rm{ln}(L_p/L_m)\right)^2}{2 \Delta^2}\right]  \, ,
\label{lpeak_given_m}
\end{equation}
where $L_m$ denotes the mean peak luminosity for a halo of mass $M$. We
adopt the scaling mentioned above, shown by the solid line in
Fig. \ref{peakl_v_mass}, $L_m = 6.78 \times 10^{10} L_\odot
\left(M/10^{12} M_\odot \right)^{4/3}$. The quantity $\Delta$ is the
dispersion in the natural logarithm of the peak luminosity, $\Delta =
\sigma_{\rm L_p}/L_p \sim 0.8$ ($0.35$ dex, in good agreement with
that observed by e.g.\ Marconi \&\ Hunt 2003).

\subsection{The Quasar Light Curve}
\label{l_curve}

The next ingredients in our theoretical modeling are the quasar light
curve and lifetime, which determine the amount of time that quasars
with a given peak luminosity spend in different intervals in
instantaneous luminosity. This represents the main difference between
our modeling and previous work: in our picture, quasars spend an
extended amount of time radiating at less than their peak luminosity,
in contrast to models in which sources follow simplified `on/off'
(`light bulb') or pure exponential light curves.  The observed quasar
luminosity function should then be thought of as a convolution of the
quasar light curve with an {\em intrinsic} distribution of sources of
a given peak luminosity. We will follow Hopkins et al. (2005c,e) and
extract the peak luminosity distribution from the observed luminosity
function, using the quasar light curves obtained from our numerical
simulations. Given the peak luminosity distribution, and the
correlation between peak luminosity and halo mass described above, we
can then predict the clustering properties of quasars.

We proceed by describing the relation between the quasar light
curve, the peak luminosity distribution, and the quasar luminosity
function.  Specifically, the quasar luminosity function can be written
as (Hopkins 2005e):
\begin{equation}
L \frac{d\Phi}{dL} (t_0) = L \frac{d}{dL} \left[ \int_{t_0 - t_q(L|L_p)}^{t_0} dt^\prime \int 
\frac{dL_p}{L_p} L_p \frac{d\dot{n}_p(t^\prime)}{dL_p} \right] \, ,
\label{lum_def}
\end{equation}
where $d\Phi(t_0)/dL$ denotes the luminosity function at
(cosmic) time $t_0$, $t_q(L|L_p)$ represents the amount of time quasars
with a given peak luminosity, $L_p$, spend at lower
luminosities, $L$, and $d\dot{n}_p/dL_p$ is the rate at which
quasars of a given $L_p$ are created or `activated'. If we further assume
that $d\dot{n}_p/dL_p$ is approximately constant over the lifetime of
the quasar activity (as shown in Hopkins et al.\ 2005e), it follows
that
\begin{eqnarray}
\label{lum_invert}
L \frac{d\Phi}{dL}(t_0) \sim \int  \frac{dL_p}{L_p} L_p \frac{d\dot{n}_p(t_0)}{dL_p} L \frac{dt_q(L|L_p)}{dL} 
\\ \nonumber
 = \int \frac{dL_p}{L_p} L_p \frac{dn_p(t_0)}{dL_p} L \frac{dP(L|L_p)}{dL} \, .
\end{eqnarray}

The second equality in the above equation further asserts that the
product of the rate of producing quasars and the amount of time
they spend at a given luminosity, is equal to the abundance
of sources multiplied by the probability of finding an object at a
given luminosity. This is justified by assuming that each quasar
has a similar activation timescale, $t_{\rm ac}$, so that
$\dot{n}_p \sim n_p/t_{\rm ac}$. The product of the lifetime, and the
rate of producing quasars, $dt_q(L|L_p)/dL \times
d\dot{n}_p/dL$, is then equal to $dP(L|L_p)/dL \times dn_p/dL$, where
the probability, $dP(L|L_p)/dL$, is the ratio of the lifetime to the
activation timescale. Essentially, this is just a refinement of the
commonly adopted proportionality between the probability of observing
an object and its lifetime.  We measure the quasar lifetime,
$dt_q(L|L_p)/dL$, directly in our simulations, and so we know the
above probability distribution up to a proportionality constant set by
the activation timescale. In this paper we will not try to predict
this proportionality constant theoretically, and hence our constraints
come solely from the {\em shape} of the luminosity function, and not
its absolute {\em normalization}. We then adopt the power law fitting
formula for the quasar lifetime from Hopkins et al. (2005b):
\begin{equation}
L_{\rm bol} \frac{dP\left(L_{\rm bol} | L_{\rm p, bol} \right)}{dL_{\rm bol}} \propto |\alpha| 
\left(\frac{L_{\rm bol}}{10^{9} L_\odot} \right)^\alpha \, .
\label{light_curve}
\end{equation}
This applies for luminosities less than the peak luminosity, $L_{\rm
bol} < L_{\rm p, bol}$; it is zero otherwise.  In this equation,
$\alpha$ is a function of the peak bolometric luminosity, $L_{\rm p,
bol}$.  Specifically, Hopkins et al. (2005b) give $\alpha = \rm{min}
\left[-0.2, -0.95 + 0.32\,log_{10}(L_{\rm p, bol}/10^{12} L_\odot)
\right]$. The above expressions are then converted from bolometric
luminosity to optical B-band luminosity using the Marconi et al.
(2004) formula, including a Jacobian factor to convert between the two
differential probability distributions.\footnote{We neglect here the
obscuration effects modeled by Hopkins et al. (2005b,d,e) which relate
the quasar light curves at different frequencies. We estimate that
these effects are less important than uncertainties in the
distribution of peak luminosities resulting from the poorly
constrained faint end of the quasar luminosity function.}
    
We then invert Eq. \ref{lum_invert} to find the distribution of quasar
peak luminosities, $L_p dn_{\rm p}/dL_p$ (Hopkins et al. 2005c,e). In
performing this inversion, we use the double power law fit from Boyle
et al. (2000), to represent the observed luminosity
function.\footnote{Specifically, we use the pure luminosity evolution
fit in which the break magnitude is a quadratic function of redshift
(Boyle et al. 2000).} We refer the reader to Hopkins et al. (2005c)
for a plot of the resulting distribution, and present here only a
qualitative description as follows. At high peak luminosity, the shape
of the peak luminosity distribution tracks the shape of the observed
luminosity function, it then reaches a peak near the break in the
observed luminosity function, and turns over at low peak
luminosity. The behavior of the distribution of peak luminosities
simply reflects the fact that quasars with large peak luminosity spend
long periods of time at low luminosity, and account, mostly by themselves,
for the faint end of the quasar luminosity function: there is no need
for quasars with small peak luminosity. The sharpness of the turnover
in the peak luminosity distribution depends, however, on the poorly
measured faint end of the quasar luminosity function. This behavior is
insensitive to which of the many measured quasar luminosity functions
we adopt, as demonstrated in Hopkins et al.\ (2005e).

\section{Which dark matter halos host active quasars?}
\label{hosts}
\nobreak

\nobreak
We now proceed to connect quasar properties with the properties of
their host dark matter halos. Our motivation here is that the
abundance and clustering of dark matter halos is well understood, and
these quantities are easily extracted from numerical simulations of
structure formation.  Moreover, the results of detailed simulations
(e.g. Springel et al. 2005c) agree well with analytic models based on
an excursion set formalism in the context of an ellipsoidal collapse
model (Sheth \& Tormen 2002). We can use these analytic models to
specify the abundance and clustering of dark matter halos, and relate
these to quasar properties. In this section we perform calculations at
$z = 2$ as an illustrative example; we generalize to other redshifts in 
\S \ref{zev}.

\begin{figure}[t]
\vbox{ \centerline{
\plotone{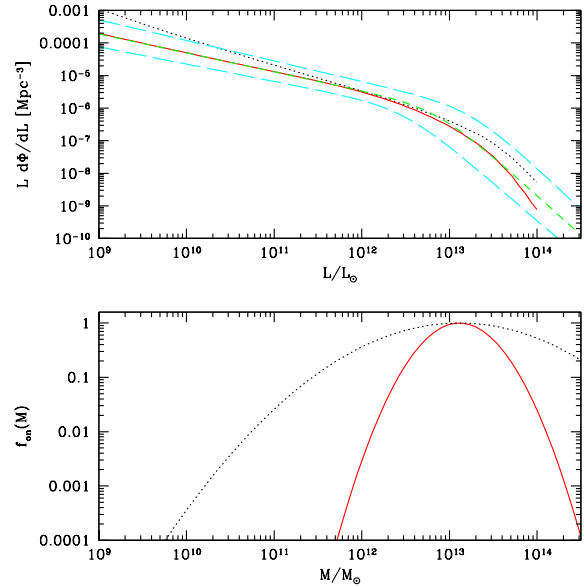}}
\caption[f2.eps]{An indication of which dark matter halos house active quasars
at $z=2$. The top panel shows the Boyle et al. (2000) fit to the $z=2$ quasar luminosity
function, as well as two model predictions. The green dashed line represents the best fit
double power law description of the data, and the cyan long-dashed lines
indicate the allowed 1-$\sigma$ range. The solid red line and black dotted lines show
model predictions. The bottom panel gives the corresponding
fraction of halos of mass M that host active quasars at $z=2$. The red solid line
is favored by the data, and the quasar light curves from the simulations. The black 
dotted line shows a model that is a less good, but still gives
an acceptable fit to the observations.   
The y-axis in the bottom panel is not normalized.
\label{fon}}  }
\end{figure} 

In order to carry out this procedure, we adopt a simple
phenomenological model: we assume that a fraction, $f_{\rm on}(M)$, of
halos of mass $M$ host active quasars.  In the future, we will attempt
to predict this quantity theoretically, but for now we will 
determine $f_{\rm on}(M)$ from the distribution of peak luminosities
above, the correlation between peak luminosity and halo mass from our
simulations, and theoretical models for the abundance of dark matter
halos.  In what follows, we will use the mass function -- i.e., the
abundance of dark matter halos with mass between $M$ and $M + dM$ --
derived by Sheth \& Tormen (2002), $dN_{\rm st}/dM$.\footnote{We
perform all calculations with the Eisenstein \& Hu (1999) fitting
formula for the transfer function, and adopt a cosmological model with
the parameters $\Omega_m=0.3$, $\Omega_\Lambda=0.7$, $\Omega_b h^2 =
0.02$, and $\sigma_8(z=0) = 0.9$.}  The peak luminosity distribution
is related to the halo mass function in our simple model by the
relation
\begin{equation}
L_p \frac{dn_{\rm p}}{dL_p} = \int dM \frac{dN_{\rm st}}{dM} f_{\rm on}(M) 
L_p \frac{dP(L_p|M)}{dL_p} \, .
\label{peak_dist}
\end{equation}
In principle, this equation can be inverted to find $f_{\rm on}(M)$. In
practice, we instead adopt a lognormal form for $f_{\rm on}(M)$,
$f_{\rm on} (M) = (\sqrt{2 \pi} \Delta_m)^{-1}
\rm{Exp}\left[-\rm{ln}(M/M_m)^2/(2 \Delta_m^2) \right]$, and determine
the parameters of the lognormal, $\Delta_m$, $M_m$, that best match
the distribution of peak luminosities, which is in turn constrained to
match the observed luminosity function (see Eq. \ref{lum_invert}). The 
lognormal form is a convenient parameterization, since, as we will
illustrate, our model predicts that quasar host halos have a well-defined
characteristic mass. The peak of the lognormal distribution conveniently
represents this characteristic mass, while the width of the distribution
indicates how broad a range of halo masses host active quasars. As
described in \S \ref{lum_mass}, we infer the peak luminosity
distribution only up to an overall normalization constant, and so we
are fitting only the shape of the $f_{\rm on}(M)$ distribution, and
not the absolute normalization.

The results of this exercise are shown in Fig. \ref{fon}. The bottom
panel of the figure shows two models for the fraction of halos of mass
$M$ that host active quasars, while the top panel shows the
corresponding predictions for the quasar luminosity function. The red
solid line indicates the best fit, while the black dotted line
describes a model that is an acceptable, but less good fit to the
data.  In our best fit model at $z=2$, active quasars are hosted by
dark matter halos with a characteristic mass, $M \sim 1.3 \times
10^{13} M_\odot$. Moreover, only a narrow range of halo masses appear
to house active quasars: the fractional width of the distribution
$f_{\rm on}(M)$ is only $\Delta_m \sim 0.75$. The narrowness of $f_{\rm
on}(M)$ is a reflection of the narrowness of the peak luminosity
distribution seen in Hopkins et al. (2005c).

On the other hand, current measurements of the quasar luminosity
function do in principle 
accommodate a substantially broader $f_{\rm on}(M)$
distribution. This is illustrated by the black dotted line in each
panel of the figure. Here, the model luminosity function is normalized to
match the observed luminosity function at a luminosity close to the break
in the luminosity function.
In this case, the characteristic mass of quasar
host halos is the same as in our best fit model, while the
fractional width of the distribution is $\Delta_m \sim 1.8$, nearly two and a half
times as large as in our best fit model. This model is clearly a worse
match to the Boyle et al. (2000) luminosity function, but it is still
within the range allowed by the measurement errors. Finally, we note that
if the mean peak luminosity, $L_m$, of Eq. \ref{lpeak_given_m} instead
follows the Wyithe \& Loeb (2003) relation, $L_m \propto M^{5/3}$, the
characteristic halo mass would be $\sim 7.5 \times 10^{12} M_\odot$, and the width
of the distribution would be $\Delta_m \sim 0.5$.

To reiterate, quasars with large peak luminosity sit in massive halos,
and spend a significant amount of time at lower luminosities. These
sources already account for the faint end of the luminosity function,
and hence one {\em over-produces the abundance of faint quasars} if
low mass halos host active quasars.  Moreover, if very massive halos
house active quasars, one might over-produce the bright end of the
quasar luminosity function.  Consequently, a wide range of quasar
luminosities corresponds to only a narrow range in host halo mass.  The
extent to which this is true depends on the details of the faint end
of the quasar luminosity function, which is thus far poorly measured.

\begin{figure}[t]
\vbox{ \centerline{ 
\plotone{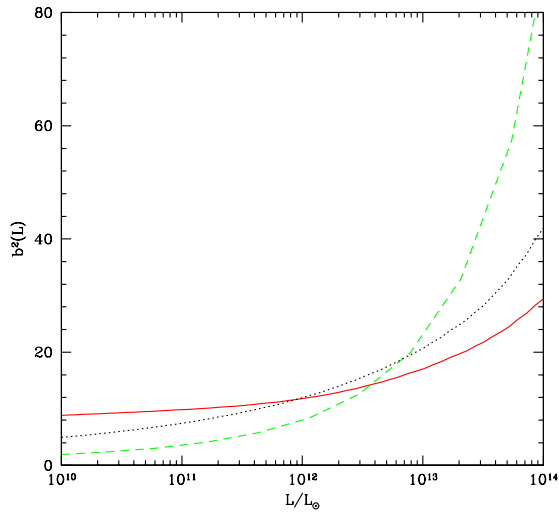}}
\caption[f3.eps]{Bias-squared of quasars as a function of their 
luminosity at $z=2$. The red solid
line shows the luminosity dependence of quasar clustering in our best-fit model (see the red solid line in Fig. \ref{fon}). The black dotted line
shows the same in our model which is a marginal fit to the quasar luminosity function (see the corresponding line in Fig. \ref{fon}). 
Finally, the green dashed line 
shows the luminosity dependence of quasar clustering in a `light bulb' model in which quasars
radiate at their peak luminosity for their entire lifetime.
\label{bias_v_lum}}  }
\end{figure} 

We can now turn our constraint on the fraction of halos of mass $M$
that host active quasars into a constraint on the luminosity
dependence of quasar clustering. It will first be useful to write down
an expression for the conditional probability that a halo of mass $M$
houses a quasar with {\em instantaneous} B-band luminosity $L$. This
is analogous to the conditional luminosity function in the
halo-occupation distribution formalism, considered in the context of
the abundance and clustering of galaxies (e.g., Yang, Mo, \& van den
Bosch 2003).  This probability distribution can be written as
\begin{equation}
L \frac{dP(L|M)}{dL} = f_{\rm on}(M) \int \frac{dL_p}{L_p} L \frac{dP(L|L_p)}{dL} L_p \frac{dP(L_p|M)}{dL_p} \, .
\label{cond_prob}
\end{equation}
We can then express the luminosity function as an integral over halo
mass. (This is in contrast to Eq. \ref{lum_invert} where we expressed
the quasar luminosity function as an integral over peak luminosity.)
This relation is
\begin{equation}
L \frac{d\Phi}{dL} = \int dM \frac{dN_{\rm st}}{dM} L \frac{dP(L|M)}{dL} \, .
\label{qso_lum_mass}
\end{equation}
An expression for the bias of a quasar of instantaneous B-band
luminosity, $L$, then follows in terms of the bias of a halo of mass
$M$,\footnote{The bias is defined so that, on large scales, the power
spectrum of halos of mass $M$ is related to the dark matter power
spectrum via $P_M (k) = b^2 P_{\rm dm}(k)$. Here, $P_M(k)$
and $P_{\rm dm}(k)$ represent the halo and dark matter power spectra,
respectively, at wavenumber $k$. We caution that this description
applies only when the scale under consideration is much larger than the
virial radius of the host halos (e.g., Scoccimarro et al. 2001.) Some
care is therefore required when connecting observed correlation
lengths with the bias of the source host halos.}
\begin{equation}
b(L) = \left[L \frac{d\Phi}{dL} \right]^{-1} \int dM \frac{dN_{\rm st}}{dM} b(M) L \frac{dP(L|M)}{dL} \, .
\label{bias_lum}
\end{equation}
To complete the calculation, we require an expression for the bias of
a halo of mass $M$, $b(M)$. We use the formula from Sheth, Mo, \&
Tormen (2001), which agrees well with measurements from N-body
simulations.

We investigate the luminosity dependence of quasar clustering in three
different models.  The first two are each based on our simulated
quasar light curves.  These two correspond to the curves shown in
Fig. \ref{fon} and are meant to bracket the possible range allowed by
present luminosity function measurements. We contrast these models
with one in which the {\em instantaneous} luminosity is related to the
halo mass in the same way that the {\em peak} luminosity is related to
the halo mass in our model, i.e., $L \propto M^{4/3}$.  Furthermore, in this scenario 
we neglect any scatter in the relation between luminosity and halo mass. This is
equivalent to the `light bulb' model assumption in which quasars
radiate at exactly their peak luminosity for their entire lifetime.

The results of this calculation are shown in Fig. \ref{bias_v_lum},
which illustrates the main point of this paper.  The plot reveals
significant differences between the three models. Our first model
corresponds to the $f_{\rm on}(M)$ distribution which best matches
the observed luminosity function, as shown in Fig. \ref{fon}. The 
bias in this model,
denoted by the red solid line, is quite flat as a function of
luminosity, before ramping up at very high luminosity. Our second model
corresponds to the broader $f_{\rm on}(M)$ distribution of Fig. \ref{fon}, 
which provides a marginal match to the observed luminosity function.
The bias in this model, denoted by the black dotted line, is qualitatively
similar to that in our best fit model, although it is a less flat function
of luminosity. Finally, in the `light bulb' model, the results are 
quite different: the quasar bias increases much more sharply with luminosity.

The difference between the bias in these models is easy to understand.
In the case of the `light bulb' scenario, there is a one-to-one
relation between luminosity and halo mass.  The range of luminosities
shown in the plot thus corresponds to a wide range in halo mass. The
quasar bias, which depends strongly on halo mass, increases sharply
with luminosity.  Considering a pure exponential light curve gives an
essentially identical result, as the implied peak luminosity
distribution has nearly the same shape (Hopkins et al.\ 2005c).  In
the case of our models based on realistic quasar light curves,
however, the entire range in luminosity corresponds to only a
relatively small range in halo mass.  As a result, the variation of quasar
bias with luminosity is rather weak. The red solid line and the black
dotted line illustrate that exactly how weak the trend with luminosity
is depends on how narrow the $f_{\rm on}(M)$ distribution is, which is
relatively poorly constrained by current luminosity function
measurements.

One might wonder about the implications of these results for attempts
to infer the lifetimes of quasars from their clustering properties (Martini \&
Weinberg 2001; Haiman \& Hui 2001). The usual view is that quasar
clustering makes it possible to distinguish whether quasars are
numerous yet short-lived, or are rare but long-lived sources. Here we
merely echo the sentiment of Adelberger \& Steidel (2005): quasar
lifetimes depend strongly on the {\em instantaneous} luminosity of
quasar activity and the `duty cycle' is larger for faint objects than
bright ones.  By `duty cycle', we mean the ratio of the abundance of
quasars in a given luminosity range to the abundance of the dark
matter halos that host them (Martini \& Weinberg 2001; Haiman \& Hui
2001; Adelberger \& Steidel 2005).  Indeed, let us consider our model
in which the characteristic halo mass at $z = 2$ is $\sim 1.3 \times
10^{13} M_\odot$, and the dispersion in halo mass is $\Delta_m \sim
0.75$. In this case, the duty cycle for quasars with instantaneous
B-band luminosity in the range $10^{10} - 10^{11} L_\odot$ is quite
large, $\sim 0.3$. On the other hand, quasars with luminosity in the
range $10^{13} - 10^{14} L_\odot$ have a duty cycle of only $\sim 8
\times 10^{-4}$.  In other words, the quasar lifetime derived from
quasar clustering should depend strongly on the luminosity of the
sources considered, and should not be interpreted as an {\em
intrinsic} lifetime (Hopkins et al. 2005a,e).

Finally, we note that our model predicts only the relative duty cycles
of faint and bright quasars, since we do not attempt to predict the
{\em absolute normalization} of the peak luminosity distribution. 
Although the bias as a function of luminosity is independent of this 
normalization, future work, incorporating theoretical estimates of the merger rates
of gas rich galaxies, will be necessary to test whether these models
can produce the large faint-quasar duty cycle implied by our best fit
case. Alternatively, the $f_{\rm on}(M)$ distribution may be less
narrow than assumed above.

\section{Redshift Evolution}
\label{zev}

It is also interesting to consider the redshift evolution of quasar
clustering.  The luminosity dependence of quasar clustering is poorly
determined at each redshift we consider (although see Croom et
al. 2005; Adelberger \& Steidel 2005), and hence measurements
(integrated over all luminosities) at different redshifts do {\em not}
currently provide a strong test of our contention that bright and
faint quasars reside in similar mass host halos.  However, it does
provide a consistency test regarding our assertion that the observed
break in the quasar luminosity function corresponds to a turnover in
the peak luminosity distribution, and of our assumed correlation
between peak luminosity and halo mass. Moreover, from Fig. \ref{fon},
we expect that quasar host halos may have a well defined
characteristic mass. It is natural then to ask how this characteristic
mass evolves with redshift (e.g., Porciani et al. 2004; Croom et
al. 2005, Wyithe \& Loeb 2005).

There are several factors that might drive redshift evolution in the
characteristic mass of quasar host halos.  In scenarios in which black
hole growth is self-regulated by feedback, the final black hole mass
is partly set by the depth of the gravitational potential well of the
host halo. In this case, one expects the normalization of the
relation between black hole mass and circular velocity to remain
constant with redshift.  The same is not true, however, for the
relation between {\em halo mass} and black hole mass: high redshift
halos of a given mass have deeper gravitational potential wells and
can grow larger black holes than halos of the same mass at lower
redshift (e.g. Wyithe \& Loeb 2003).  More specifically, from the
scaling $M_{\rm bh} \propto v_c^4$, and connecting circular velocity
to halo mass, we have $M_{\rm bh} \propto
\left[\zeta(z)\right]^{2/3} (1+z)^{2} M^{4/3}$ (Wyithe \& Loeb 2003), where $\zeta(z) =
\left[\Omega_m/\Omega_m(z)\right] \left[\Delta_c/18 \pi^2\right]$, and
$\Delta_c = 18 \pi^2 + 82 \left(\Omega_m(z) - 1 \right) - 39
\left(\Omega_m(z) - 1 \right)^2$. Here, $\Omega_m(z)$ denotes the
matter density at redshift $z$ in units of the critical density, and
$\Delta_c$ is the collapse overdensity according to the fitting
formula of Bryan \& Norman (1998).  We further assume that the
relation between peak luminosity and final black hole mass does not
evolve with redshift, in which case the redshift evolution of the peak
luminosity-halo mass relation is the same as that of the black hole
mass-halo mass relation.  These assumptions are further confirmed in
numerical simulations at different redshifts, or more accurately, simulations
in which we vary the properties of the merging galaxies to mimic redshift
evolution (Robertson et al.\ 2005).  Moreover, the break quasar luminosity, $L_\star$, evolves
with redshift (e.g. Boyle et al. 2000), as does the shape of the halo
mass function.

We aim, then, to piece together each of these evolving ingredients,
and determine the evolution of quasar clustering with redshift,
employing the same methodology as in \S \ref{lum_mass} and \S
\ref{hosts}. To this end, we adopt the pure luminosity evolution (PLE)
double power law model of Boyle et al. (2000) in which the break
magnitude varies quadratically with redshift, $M_\star(z) = -22.65 +
1.35 z - 0.27 z^2$. For simplicity, we will further assume that the
{\em scatter} in the relation between peak quasar luminosity and halo
mass is independent of redshift.  We can then infer the distribution
of quasar peak luminosities (Eq. \ref{lum_invert}) and determine which
dark matter halos host active quasars ($f_{\rm on}(M)$ in
Eq. \ref{peak_dist}) at each of several redshifts. At each redshift,
we try to determine the $f_{\rm on}(M)$ that provides the best fit to
the quasar luminosity function, noting that current luminosity
function measurements tolerate a wide range of values for $f_{\rm
on}(M)$, as illustrated in Fig. \ref{fon}.  We expect the qualitative
feature illustrated in the previous section, that quasar bias depends
weakly on luminosity, to hold at all redshifts. In order to illustrate
the redshift evolution of the typical quasar bias, we then consider a
luminosity-averaged quasar bias defined by
\begin{equation}
\bar{b} = \left[\int_{L_{\rm min}}^{L_{\rm max}} dL \frac{d\Phi}{dL} \right]^{-1} 
\int_{L_{\rm min}}^{L_{\rm max}} dL\,b(L) \frac{d\Phi}{dL} \, .
\label{bbar}
\end{equation} 
In what follows, we choose $L_{\rm min} = 0.1 L_\star(z)$, and $L_{\rm
max} = 10 L_\star(z)$.  Although this does not correspond precisely to
the quasar bias that is measured observationally (e.g., Croom et
al. 2005), our results are not very sensitive to our choices for
$L_{\rm min}$ and $L_{\rm max}$.  After all, the bias in our model
depends only weakly on luminosity.

\begin{figure}[t]
\vbox{ \centerline{
\plotone{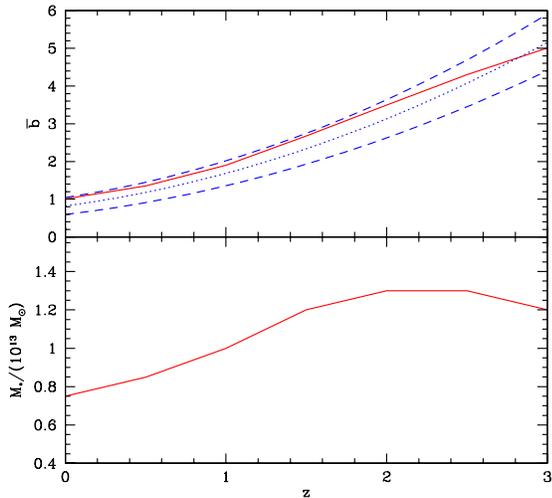}}
\caption[f4.eps]{Quasar bias as a function of redshift. The top panel shows, as a red solid line, our 
theoretical
expectation for the redshift evolution of quasar bias. The kink in the
model prediction is likely an artifact of extending the PLE model of Boyle
et al. (2000) beyond the redshift extent of their measurement.
 The blue dotted line shows the best fit measurement
from Croom et al. (2005), while the blue dashed-lines indicate the allowed 1-$\sigma$ range implied
by their measurement. The bottom panel shows the corresponding characteristic mass of 
dark matter halos that host active quasars as a function of redshift.
\label{bofz}}  }
\end{figure}

The results of this calculation are shown in Fig. \ref{bofz}. The
bottom panel of the figure shows the characteristic mass of quasar
host halos -- specifically, the center of the lognormal distribution
$f_{\rm on} (M)$ (Eq. \ref{peak_dist}) -- for several redshifts. The
figure clearly illustrates that the characteristic mass of quasar host
halos evolves relatively little with redshift in our model. The reason
for this is as follows. The characteristic mass of quasar host halos
is primarily determined, within the context of our model, by the
turnover in the distribution of quasar peak luminosities and the
correlation between peak luminosity and halo mass. The turnover in the
distribution of quasar peak luminosities is in turn set by the
position of the break in the quasar luminosity function. The redshift
evolution of the position of the break in the quasar luminosity
function is, however, compensated by evolution in the peak
luminosity-halo mass relation, and somewhat by changes in the shape of
the halo mass function. Consequently, we find that our model produces
the observed evolution in the break luminosity {\em at close to fixed
host halo mass}: the characteristic host halo mass appears to vary by
less than a factor of $\sim 2$ between $z = 0$ and $z = 3$. We note
that the figure shows a turnover in the host halo mass near $z \sim
2.5$, which corresponds to a similar turnover in $L_\star(z)$, but
this is likely an artifact of extending the PLE model of Boyle et
al. (2000) beyond the redshift extent of their measurement ($z \sim
2.3$). Finally, we remark that if the peak luminosity scales with
halo mass as $L_p \propto M^{5/3}$ then the results are qualitatively similar,
but the characteristic halo masses are a factor of $\sim 1.5 - 2$ times 
smaller.

The redshift evolution of the break luminosity is then a reflection of
the self-regulated nature of quasar activity: halos that host quasars
have deeper potential wells at high redshift and can thereby grow
larger, more luminous black holes at high redshift than at low
redshift. In this sense, black hole growth is {\em anti-hierarchical}:
massive black holes form at higher redshifts than low-mass black
holes (Cowie et al. 2003). In order to understand the physics that drives
this anti-hierarchical growth in more detail, we need to understand what
sets the characteristic mass of quasar host halos, a topic we briefly
speculate on in the concluding section.

We show the resulting prediction for quasar bias as a function of
redshift in the top panel of Fig. \ref{bofz}. We compare our
theoretical predictions with the quasar bias measured by Croom et
al. (2005).\footnote{Croom et al. (2005) derive the quasar bias
assuming a slightly different cosmological model than we adopt
here. The difference between the bias in our two models should,
however, be small compared to the statistical errors in the
measurement. In addition, the quantity Croom et al. (2005) measure is
a little different from the luminosity-averaged bias we predict in
Eq. \ref{bbar}. Again, the difference between our definitions of
quasar bias should be small compared to statistical measurement
errors.}  The figure illustrates that our theoretical predictions
agree with the measured bias as a function of redshift, although
our results are a bit higher the best fit measurements. This slight overprediction 
is somewhat sensitive to the assumed peak luminosity-halo mass relation, however, 
in the sense that assuming $L_{p}\propto M^{5/3}$ produces better agreement
with the best fit measurements. 
The key qualitative feature of the figure is that, even though quasars at $z
\sim 0$ and $z \sim 3$ reside in similar host halos, their clustering
properties differ significantly.  Specifically, quasars at $z \sim 0$
should be close to un-biased ($\bar{b} \sim 1$), while quasars at $z
\sim 3$ are highly biased, with $\bar{b} \sim 5$.  The reason for
this is simply that halos of mass 
$\sim 7.5 \times 10^{12} M_\odot - 1.5 \times 10^{13} M_\odot$
correspond to rare, high-$\sigma$ peaks at $z \sim 3$, and are thus
highly-clustered. On the other hand, the variance of the density field
smoothed on the same mass scale is close to the collapse threshold at
$z \sim 0$, $\sigma(M) \sim \delta_c$, and hence these halos
faithfully trace the matter distribution (see also Croom et al. 2005, Wyithe \& Loeb 2005)
near $z \sim 0$.  This trend is to be expected in the context of our 
model if mergers involving gas-rich galaxies occur mainly in dense
environments at high redshifts, but in more isolated regions at
$z\sim 0$.

\section{Conclusion}
\label{conclusion}

In this paper, we have connected the properties of quasars, as
determined from numerical simulations of galaxy mergers
(Springel et al. 2005a, Di Matteo et al. 2005, Robertson et al. 2005), to the properties of 
the dark matter halos
that host them. We find that bright and faint quasars reside in
similar mass halos with characteristic masses close to $\sim 1 \times
10^{13} M_\odot$. As a result, we predict that quasar clustering
should depend only weakly on quasar luminosity. Furthermore, we
predict that the characteristic mass of quasar host halos should
evolve only weakly with redshift. We note that Di Matteo et al. (2003)
also found, using cosmological simulations with a simple model for black hole
growth, that quasars at low and high redshift reside in similar mass host halos,
although they associate quasars with less massive halos than we find presently. 
Our conclusions invite two obvious
questions. The first question is of a theoretical nature: what physics
sets the characteristic mass scale of quasar host halos? The second
question is an observational one: what do {\em observational
measurements} of the luminosity dependence of quasar clustering
find?

We will largely defer answering the first question to future work, but
a plausible explanation is that the most luminous quasars at a given
redshift are triggered by the most massive gas rich galaxies merging
at that time.  The halos that host these objects will be determined by
the evolution of merger rates, depending on environment, and the gas
content of the galaxies they contain.  We note that more precise
luminosity function measurements will be helpful in obtaining tighter
constraints on the width of the $L_p$ distribution, and on the mass
range of halos that host active quasars. Observations of e.g.\ the 
Eddington ratio distribution and active black hole mass function 
can further constrain this distribution 
at faint luminosities where the observed luminosity function 
provides only weak limits (Hopkins et al.\ 2005e). 
Likewise, the distribution of host masses can be constrained by 
observations of the quasar host galaxy luminosity function; these  
find an approximately lognormal distribution with narrow 
width $\Delta_{L}\sim\Delta_{m}=0.5$
($\sim0.6-0.7$ magnitudes) and a peak corresponding to 
the stellar mass of quasar hosts with $L_{p} \sim L_{\star}$
(Bahcall et al.\ 1997; McLure et al.\ 1999; Hamilton et al.\ 2002), 
close to that predicted by our best-fit model. 
Further progress can be made
theoretically with more detailed semi-analytic calculations
incorporating galaxy merger rates (Kauffman \& Haehnelt 2002), or with
cosmological simulations incorporating black hole growth and feedback.
 
We now address the second question.  There have been two recent
observational attempts to measure the luminosity dependence of quasar
clustering.  First, Croom et al. (2005) examine the luminosity
dependence of quasar clustering from $\sim 20,000$ 2dF quasars,
finding {\em no evidence} for any luminosity dependence. Their
measurement, however, spans only a factor of $\lesssim 20$ in
luminosity.  Second, Adelberger \& Steidel (2005) examine the
luminosity dependence of quasar clustering using the galaxy-AGN
cross-correlation function, rather than the quasar auto-correlation
function.  This approach, initially advocated by Kauffmann \& Haehnelt
(2002), takes advantage of the fact that galaxies are much more
abundant than quasars: therefore, statistical measurements of the
galaxy-AGN cross correlation function are correspondingly tighter than
measurements of the quasar auto-correlation function. Furthermore,
Adelberger \& Steidel (2005) probe the luminosity dependence of quasar
clustering with a much larger dynamic range, roughly a factor of $10$
in magnitude, or a factor of $10^4$ in luminosity. Their result is,
again, that quasar clustering is independent of
luminosity. Specifically, they find that the correlation length for
quasar sources with $-30 < M_{1350} < -25$ is $r_0 = 4.7 \pm 2.3$, and
$r_0 = 5.4 \pm 1.2$ for sources with $-25 < M_{1350} < -19$, where
$M_{1350}$ denotes an AB magnitude at a rest-frame wavelength of
$1350$\AA. The statistical precision of these results is not high,
but again, they are qualitatively consistent with our picture. We
eagerly anticipate more precise measurements from SDSS and 2dF in
the near future, which should provide a more definitive test of this
picture for quasar formation and evolution.

Finally, we remark that in this paper we confined our theoretical calculations
to large scales where linear biasing is an accurate description of clustering,
but it would be interesting to extend calculations to smaller scales. Indeed,
Hennawi et al. (2005) find that the quasar correlation function is an order
of magnitude larger, at proper separations of $r \lesssim 40$ kpc/h, than expected
based on extrapolating clustering measurements from large scales. This is likely
evidence that quasar clustering is associated with galaxy mergers in dense environments,
but this should be quantified.   

\acknowledgments  We are extremely grateful to Tiziana Di Matteo and 
Volker Springel
for their role in the simulations which inspired this analysis, and for 
helpful comments on a draft.
This work was supported in part by NSF grants ACI 
96-19019, AST 00-71019, AST 02-06299, and AST 03-07690, and NASA ATP
grants NAG5-12140, NAG5-13292, and NAG5-13381.  The simulations used
in the analysis were performed at the Center for Parallel
Astrophysical Computing at the Harvard-Smithsonian Center for
Astrophysics.

\end{document}